\global\def\draftcontrol{0}

   \def\versionno{ effective action of flux compactifications -- draft   }

\catcode`\@=11

\expandafter\ifx\csname draftcontrol\endcsname\relax\global\def\draftcontrol{0}
\fi

{\count255=\time\divide\count255 by 60
\xdef\hourmin{\number\count255}
\multiply\count255 by-60\advance\count255 by\time
\xdef\hourmin{\hourmin:\ifnum\count255<10 0\fi\the\count255}}
\def\draftdate{\number\month/\number\day/\number\year\ \ \ \hourmin }

\newcommand\makepapertitle{\par
  \begingroup
    \renewcommand\thefootnote{\@fnsymbol\c@footnote}%
    \def\@makefnmark{\rlap{\@textsuperscript{\normalfont\@thefnmark}}}%
    \long\def\@makefntext##1{\parindent 1em\noindent
            \hb@xt@1.8em{%
                \hss\@textsuperscript{\normalfont\@thefnmark}}##1}%
     \newpage
     \global\@topnum\z@   
     \@makepapertitle
     \thispagestyle{empty}\@thanks
  \endgroup
  \setcounter{footnote}{0}%
  \global\let\thanks\relax
  \global\let\makepapertitle\relax
  \global\let\@makepapertitle\relax
  \global\let\@thanks\@empty
  \global\let\@author\@empty
  \global\let\@date\@empty
  \global\let\@title\@empty
  \global\let\title\relax
  \global\let\author\relax
  \global\let\date\relax
  \global\let\and\relax
  \def\version{\let\version\@version\@gobble}
}
\def\@makepapertitle{%
  \newpage
   \ifnum\draftcontrol=1 {}
   \version\versionno
   \vskip 3em%
   \else
   \hfill\hbox to 3cm {\parbox{4cm}{\@pubnum}\hss}%
   \vskip 3em%
   \fi
   \begin{center}%
   \let \footnote \thanks
     {\LARGE {\@title}}%
     \vskip 1.5em%
     {\normalsize
       \lineskip .5em%
       \begin{tabular}[t]{c}%
         \@author
       \end{tabular}\par}%
     \vskip 1.5em%
     {\@bstract}%
     \end{center}%
     \vskip 1.5em
     \@date%
   \par
}

\gdef\@pubnum{}
\def\pubnum#1{%
  \gdef\@pubnum{#1}}

\gdef\@bstract{}
\def\Abstract#1{%
  \gdef\@bstract{%
   \parbox{\textwidth-0pc}{%
   \centerline{\bf Abstract}\penalty1000%
\kern.2cm%
\noindent
\renewcommand\baselinestretch{1.0}%
{#1}}}
}

\def\ps@paper{\let\@mkboth\@gobbletwo%
     \ifnum\draftcontrol=1
	\def\@oddfoot{\hbox to \textwidth{\tiny \versionno \hfil\tiny\draftdate}%
	\hskip -\textwidth \hbox to \textwidth{\hfil\rm\thepage\hfil}}%
     \else\def\@oddfoot{\hbox to \textwidth{\hfil\rm\thepage\hfil}}
     \fi
     \let\@evenfoot\@oddfoot
}

\def\body{\clearpage
          \pagestyle{paper}
	}

\def\@version#1{\ifnum\draftcontrol=1
\typeout{}\typeout{#1}\typeout{}
\vskip3mm\centerline{\hbox{\fbox{\normalsize{\tt DRAFT -- #1 -- }
                   {\draftdate}}}}\vskip3mm
\fi}
\let\version\@version
\long\def\eqlabel#1{\ifnum\draftcontrol=1
                    \tag@false  
                    \tag*{(\theequation) \hbox to -0.2cm{\hspace{0cm}\small{#1}\hss}}
                    \refstepcounter{equation}
                    \edef\@currentlabel{\theequation}
                    \ltx@label{#1}          
                    \else
                    \label{#1}
                    \fi
                    }
\let\st@bibitem\@bibitem
\let\st@lbibitem\@lbibitem
\ifnum\draftcontrol=1
  \def\@bibitem#1{%
    \st@bibitem{#1}\a@@label{#1}\ignorespaces}
  \def\@lbibitem[#1]#2{%
    \st@lbibitem[#1]{#2}\a@@label{#2}\ignorespaces}
  \def\a@@label#1{%
    \gdef\a@lab{\smash{\normalfont\small#1}}
    \ifvmode
      \if@inlabel
        \global\setbox\@labels\hbox{%
          \llap{\a@lab\let\a@lab\relax
                \kern\@totalleftmargin\kern\marginparsep}%
          \box\@labels}%
      \fi
    \fi}
\fi

\documentclass[12pt,letterpaper]{article}

\usepackage{amsmath,amssymb,array,calc,rotating,epsfig,psfrag}
\usepackage[nosort]{cite}

\ifnum\draftcontrol=1
\fi

\renewcommand\baselinestretch{1.25}
\renewcommand\section{\@startsection {section}{1}{\z@}%
                                   {-3.5ex \@plus -1ex \@minus -.2ex}%
                                   {2.3ex \@plus.2ex}%
                                   {\normalfont\large\bfseries}}
\renewcommand\subsection{\@startsection{subsection}{2}{\z@}%
                                   {-3.25ex\@plus -1ex \@minus -.2ex}%
                                   {1.5ex \@plus .2ex}%
                                   {\normalfont\normalsize\bfseries}}
\renewcommand\subsubsection{\@startsection{subsubsection}{3}{\z@}%
                                   {-3.25ex\@plus -1ex \@minus -.2ex}%
                                   {1.5ex \@plus .2ex}%
                                   {\normalfont\normalsize\it}}
\renewcommand\paragraph{\@startsection{paragraph}{4}{\z@}%
                                   {-3.25ex\@plus -1ex \@minus -.2ex}%
                                   {1.5ex \@plus .2ex}%
                                   {\normalfont\normalsize\bf}}

\numberwithin{equation}{section}

\def\ie{{\it i.e.}}

\def\revise#1       {\raisebox{-0em}{\rule{3pt}{1em}}%
                     \marginpar{\raisebox{.5em}{\vrule width3pt\
                     \vrule width0pt height 0pt depth0.5em
                     \hbox to 0cm{\hspace{0cm}{%
                     \parbox[t]{4em}{\raggedright\footnotesize{#1}}}\hss}}}}

\newcommand\nxt[1]  {\\\fnxt#1}

\def\calb         {{\cal B}}

\def\calf         {{\cal F}}

\def\cali         {{\cal I}}

\def\calk         {{\cal K}}
\def\call         {{\cal L}}
\def\calm         {{\cal M}}

\def\calv         {{\cal V}}

\def\del          {\partial}

\def\Im           {{\rm Im\hskip0.1em}}

\def\sqr#1#2{{\vcenter{\vbox{\hrule height.#2pt
 \hbox{\vrule width.#2pt height#1pt \kern#1pt
 \vrule width.#2pt}\hrule height.#2pt}}}}

\newcommand{\ft}[2]{{\textstyle{\frac{#1}{#2}}}}

\def\om{\Omega}

\def\a{\alpha}

\def\r{\rho}

\def\dd{\delta}
\def\tg{\tilde{g}}
\def\LL{\Lambda}
\def\ww{\omega}
\def\n{\nabla}
\def\tn{\tilde{\nabla}}
\def\bG{{\bar{G}}}

\def\hg{\hat{g}}
\def\tcalm{\tilde{\calm}}
\def\bD3{\overline{D3}}

\def\si{\sigma}
\def\tcalv{\tilde{\calv}}

\catcode`\@=12

\begin{document}


\title{On effective action of string theory flux  compactifications}

\pubnum{%
hep-th/0312076}
\date{December 2003}

\author{
Alex Buchel\\[0.4cm]
\it Department of Applied Mathematics\\
\it University of Western Ontario\\
\it London, Ontario N6A 5B7, Canada\\
\it Perimeter Institute for Theoretical Physics\\
\it Waterloo, Ontario N2J 2W9, Canada\\[0.2cm]
}

\Abstract{
We discuss four dimensional effective actions of string theory 
flux compactifications. These effective actions describe 
four dimensional gravity coupled to overall K\"ahler modulus 
of the compactification manifold.   
We demonstrate the agreement 
between ten dimensional equations of motion of supergravity 
with localized branes, and equations of motion derived from the 
effective action. The agreement is lost however if one 
evaluates the full effective action on the equations of motion for 
a subset of the supergravity modes, provided these modes  depend on-shell on 
the K\"ahler modulus.
}


\makepapertitle

\body

\version\versionno

\section{Introduction}
Probably the biggest surprise of recent cosmological measurements is
the evidence that the cosmological constant of the Universe is
positive. It is thus important to understand how (and if)
four dimensional de-Sitter vacua arise in string theory
compactifications. The problem is complicated by the well known no-go
theorem in supergravity \cite{gib,mn} which forbids non-singular
(warped) de-Sitter compactifications with finite four dimensional Newton's constant
\footnote{
Having a finite four dimensional Newton's constant is a crucial qualifier, as it is
straightforward to construct explicit supergravity warped product
backgrounds $dS_4\times \calm_6$ for non-compact six-dimensional manifolds
$\calm_6$ \cite{b1,blw,b2}.}.  
This no-go theorem can be phrased in the statement on the energy condition
on the matter stress tensor upon consistent Kaluza-Klein (KK) reductions\footnote{Most recently 
this has been emphasized in \cite{tw,dealwis}.}: if the matter of the higher dimensional theory obeys the strong 
energy condition, the matter of the lower dimensional effective description 
would satisfy the strong energy condition as well, implying that lower dimensional de Sitter 
(or more generally accelerating Universe) is not possible. In ten dimensional supergravities 
matter satisfies the strong energy condition. 
  
The situation is different with positive tension extended objects, D-branes, in string theory.
In ten dimensions, whether or not a D$p$-brane satisfies the strong energy condition 
depends on its codimension. Specifically, codimension three and higher  branes ($p<7$) 
satisfy strong energy condition. In consistent supergravity KK reductions the effective 
lower dimensional action can be obtained by integrating out compact directions 
of a ten dimensional supergravity effective action. 
If one assumes that the same prescription is true for a D-brane effective action, 
one immediately concludes that upon  KK reduction D-branes can 
be sources for de Sitter.  The point simply is that KK reduction 
transverse to a brane {\it decreases} its codimension, thus making 
lower dimensional accelerating Universe possible. In explicit string construction 
this idea has been realized for the first time in \cite{kklt}, KKLT.  
 The starting point of KKLT
construction is type IIB string theory compactified on Calabi-Yau (CY)
3-folds $\calm_6$ with imaginary self-dual (ISD) 3-form
fluxes. Classically, as discussed in details in
\cite{gvw,gkp} (GKP), for a 
suitable choice of a CY and fluxes, one can fix all complex 
structure moduli (including the axion-dilaton), and get in four
dimensions {\it no-scale} models
\cite{ns1,ns2,ns3}: a vanishing four dimensional
cosmological constant and a complex K\"ahler modulus $\r$  (related to
the overall size of the compactification manifold).  
This no-scale structure is expected to be spoiled by 
string tree level $\a'$ corrections, as well as perturbative 
string loops and string instanton corrections \cite{gkp}.
Indeed,   in KKLT it was demonstrated that string theory instanton corrections \cite{w96}
modify the no-scale structure of GKP, leading to a supersymmetric
$AdS_4$ vacuum with fixed $\r$. 
Lastly, it was argued that,  for suitable choice of parameters,
anti-D3 brane would lift $AdS_4$ vacuum to a $dS_4$ background, 
with all moduli of the compactification manifold stabilized\footnote{Some general 
aspects of moduli fixing and potentials 
in string theory compactifications 
are also discussed in \cite{gid}.}. 
This is possible because $\bD3$ is a codimension 
zero defect in the effective four dimensional description, and thus 
violates the strong energy condition. 
 
An obvious question of KKLT construction is whether  effective description of 
$\bD3$ is indeed given by simple dimensional reduction of its ten dimensional 
effective action. The question is nontrivial, as KKLT model does not allow 
for a consistent ten dimensional lift: the K\"ahler modulus stabilization involves 
string instanton corrections, which are highly non-local from the 
ten dimensional perspective (assuming their ten dimensional description exists at all).
Thus strictly speaking, there is no starting point for the derivation of the 
$\bD3$ effective action in KKLT model. To the best of our knowledge, there is no consistent  derivation 
of the effective brane description in the Kaluza-Klein reduction for any compactification 
in the literature as well.

In section 2 we discuss a simple exactly soluble $D$-dimensional toy  model in which we confirm 
that contrary to a supergravity mode, in a consistent KK reduction to $d<D-2$ dimensions, 
the energy condition of the 
extended positive tension $d$-dimensional defect indeed changes. 
The defect contribution to the $d$-dimensional vacuum energy $\LL_d$ is positive. 
The apparent conflict with the energy condition upon KK reduction is resolved 
because the defect introduction 'shifts' the stabilization value of the overall K\"ahler modulus 
of the compactification manifold. This is precisely what is observed 
in KKLT construction. However, though the vacuum energy in the toy model compactification 
increases, it can never become positive. 
The latter point is just a special case of the no-go theorem for supergravity 
with localized  sources satisfying a BPS-like condition of Ref.~\cite{gkp}.  
Additionally, we explain the dangers of using incomplete set of $D$-dimensional 
equations of motion to deduce lower dimensional effective action.  

In section 3 we discuss type IIB string theory   flux compactifications.
Though we work in the supergravity approximation with 3-branes and orientifold 3-planes 
only, we expect that our results can be extended to more general F-theory 
compactifications. We demonstrate that the same prescription as the one employed 
in section 2 leads to a consistent effective description of four dimensional gravity 
coupled to the overall K\"ahler modulus of the compactification manifold.    
'Consistent' means that equations of motion derived from the effective action
are equivalent to   ten dimensional equations of motion. 
This resolves some puzzles of the consistent Kaluza-Klein compactifications 
of string theory with fluxes and localized sources raised in \cite{dealwis}.
We conclude with discussion of partially integrating out supergravity modes
from the lower dimensional effective action.

\section{Defect effective action in consistent KK reductions}
In this section\footnote{Results reported in this section 
were obtained in collaboration with Rob Myers.} we consider a soluble toy model where we verify that in consistent 
Kaluza-Klein reductions the extended object lower dimensional effective action is obtained 
by integrating out the compact coordinates of the higher dimensional effective description.

Consider the following $D=d+q$ dimensional\footnote{We assume $d>2$.} model of gravity coupled to the $q$-form field 
strength $F_{[q]}$
\begin{equation}
S_{D}=\frac{1}{2k_D^2}\int_{\calm_{D}}d^{D}\xi
\sqrt{-g}\biggl(
R^{D}-\frac {1}{2q!}F^2_{[q]} 
\biggr)\,,
\eqlabel{Daction}
\end{equation}
where $k_D$ is $D$-dimensional gravitational coupling.
We will use effective action \eqref{Daction}  
to study Kaluza-Klein reductions of direct warped product compactifications  
$\calm_{D}=\calm_d\times S^{q}$ on $S^q$,
\begin{equation}
\begin{split}
dS_{D}^2&=g_{MN}d\xi^Md\xi^N\\
&= \si^{-2 }\ \tg_{\mu\nu}dx^\mu dx^{\nu}
+\si^{\frac{2(d-2)}{q}}\ \left(dS^{q}\right)^2
\,,
\end{split}
\eqlabel{dmetric}
\end{equation}
where $\si=\si(x)$ is a $q$-sphere K\"ahler modulus, $\tg_{\mu\nu}=\tg_{\mu\nu}(x)$ is the metric 
on $\calm_d$, and $\left(dS^{q}\right)^2$ is the metric 
on a round $S^{q}$. The metric ansatz \eqref{dmetric} is supplemented with 
\begin{equation}
F_{[q]}=b\ vol(S^q)\,.
\eqlabel{flux}
\end{equation} 
Additionally, we will consider $N\gg 1$, $d$-dimensional neutral
domain walls of tension $T_d>0$, uniformly distributed on  $S^{q}$.
Each domain wall has a DBI-type action
\begin{equation}
S_d=-T_d\int_{\calm_d}\ d^dx\sqrt{-\tg}\ \si^{-d}\,. 
\eqlabel{loc}
\end{equation}
The total effective action is 
\begin{equation}
S_{tot}=S_{D}+S_{(loc)}=S_D+\sum_{i=1}^N S_d\,,
\eqlabel{stot}
\end{equation}
where the summation is over all localized defects on $S^q$.

Note that the stress tensor of the localized defects is   
\begin{equation}
T_{MN}^{(loc)}=-\frac{2}{\sqrt{-g}}\frac{\dd S_{(loc)}}{\dd g^{MN}}\,,
\eqlabel{stress}
\end{equation}
so
\begin{equation}
T^{(loc)}_{\mu\nu}=-\frac{T_d N}{ vol(S^q)\ \si^{d}}\ \tg_{\mu\nu},\qquad
T^{(loc)}_{mn}=0\,.
\eqlabel{resT}
\end{equation}
In $D$-dimensions, the stress tensor  $T_{MN}^{loc}$ satisfies the strong energy 
condition provided 
\begin{equation}
T_{tt}^{(loc)}-\frac{1}{D-2}\ g_{tt}\ T_K^{K\ (loc)}\ >\ 0\,, 
\eqlabel{encond}
\end{equation}
which using \eqref{resT} evaluates to 
\begin{equation}
-\frac{T_d N}{ vol(S^q)\ \si^{d}} \ \frac{q-2}{D-2}\ \tg_{tt}\ >\ 0\,.
\eqlabel{enc2}
\end{equation} 
Clearly, codimension higher than two  ($q>2$) defects satisfy the strong 
energy condition, while defects of lower codimension violate it.

In what follows we assume that $\si$ is constant. 
We consider first  $D$-dimensional solutions  of  \eqref{stot}, 
and show that they are equivalent to the solutions of the effective 
$d$-dimensional action, obtained by simply integrating out the $S^q$ coordinates in 
\eqref{stot}.

\subsection{$D$-dimensional equations of motion}
With the flux ansatz \eqref{flux}, the $q$-form Bianchi identity are trivially satisfied. 
We are left with Einstein equations
\begin{equation}
\begin{split}
R_{MN}=&\frac{1}{2(q-1)!}\left(F_{M\a_2\cdots\a_q}F_N\ ^{\a_2\cdots\a_n}
-\frac{q-1}{q(d+q-2)}F^2_{[q]} g_{MN}\right)\\
&+k_D^2\left(T^{loc}_{MN}-\frac {1}{d+q-2} T^{loc}g_{MN}\right)\,.
\end{split}
\eqlabel{ein}
\end{equation}
Eq.~\eqref{ein} for the components along the defect implies that $\calm_d$ 
is an Einstein manifold 
\begin{equation}
r_{\mu\nu}^{(d)}=\LL_d \tg_{\mu\nu}\,,
\end{equation}
with 
\begin{equation}
\LL_d = -\frac{b^2(q-1)}{2(d+q-2)}\si^{2(1-d)}-k_D^2\frac{T_d N}{vol(S^q)\
\si^{d}}\ \frac{q-2}{d+q-2}\,.
\eqlabel{munu}
\end{equation}
Note from \eqref{munu} that as long as defects satisfy the strong energy condition in $D$ dimensions, \ie, $q>2$,
in a consistent KK reductions they  {\it can not} produce $d$-dimensional accelerating Universe. 
As we will show in section 3, this is a special case of a more general result that follows from a no-go 
theorem for supergravity compactifications with branes satisfying BPS-like condition of \cite{gkp}.
From \eqref{munu} is appears that a positive tension defect satisfying a strong energy 
condition provides a {\it negative}
contribution to  $d$-dimensional cosmological constant. This conclusion would be correct, 
if before and after defect introduction the K\"ahler modulus of the compactification manifold 
(in this case $\si$) would not change. In fact, 
the K\"ahler modulus is shifted so that after all, the contribution of defect tension 
to $\LL_d$ is positive. Indeed, from the Einstein equations along the $S^q$ directions
\begin{equation}
(q-1)\si^{2(2-d)/q}=\frac{b^2}{2}\ \frac{d-1}{(d+q-2)}\ \si^{2(2-d)}
+k_D^2 \frac{T_d N}{ vol(S^q)\
\si^{(d-2)}}\ \frac{d}{d+q-2}\,.
\eqlabel{transf}
\end{equation}
From \eqref{transf}  it follows that defect introduction always results in the 
expansion of  $S^q$, \ie, $\si$ increases. Using \eqref{munu}, \eqref{transf},
\begin{equation}
\LL_d=-\frac{b^2}{2d}\ \frac{1}{ \si^{2d-2}}-\frac{(q-2)(q-1)}{d}\ \frac{1}{ \si^{2(d+q-2)/q}}\,,
\eqlabel{newLL}
\end{equation}
so, it is monotonically 
increasing  function of $\si$ for $q>2$. 
Thus, introduction of localized sources \eqref{loc} will always  
increase $\LL_d$.

\subsection{$d$-dimensional effective description}
We would like to reproduce the results of the previous section from the 
lower dimensional effective action, obtained by integrating out $S^q$ coordinates. 
For constant $\si$ and the metric and flux ansatze \eqref{dmetric} and \eqref{flux}  correspondingly,
we find 
\begin{equation}
\sqrt{-g}\ \biggl(R^{D}-\frac{1}{2q!}F^2_{[q]}\biggr)=
\sqrt{-\tg }\sqrt{g_{S^q}}\ \si^{-2}\biggl(r^{d}\si^2+(q-1)q\si^{2(2-d)/q}
-\frac 12 b^2 \si^{4-2d}\biggr)\,.
\eqlabel{reduc}
\end{equation}
Thus Kaluza-Klein reduction of \eqref{Daction}  yields
\begin{equation}
S_D^{eff}=\frac{vol(S^q)}{2k_D^2}\int_{\calm_d}\ d^dx\sqrt{-\tg}\biggl(r^d-\calv_{bulk}(\si)\biggr)\,,
\eqlabel{sdeff}
\end{equation}
where 
\begin{equation}
\calv_{bulk}=-(q-1)\ q\ \si^{2(2-d-q)/q}
+\frac 12 b^2 \si^{2-2d}\,.
\eqlabel{vbulk}
\end{equation}
 The KK reduction of the defect action $S_{(loc)}$ is simply 
\begin{equation}
S_{(loc)}^{eff}=-N T_d\int_{\calm_d}\ d^dx\sqrt{-\tg}\ \si^{-d}\equiv \int_{\calm_d}\ d^dx\sqrt{-\tg}\ \biggl(
-\calv_{(loc)}(\si)\biggr)\,.
\eqlabel{sloceff}
\end{equation}
Notice that the stress tensor computed from $S_{(loc)}^{eff}$ violates the strong energy condition in 
$d$-dimensions. This is so because extended defects are now codimension zero objects.   
The complete effective action reads 
\begin{equation}
\begin{split}
S^{eff}&=S^{eff}_D+S^{eff}_{(loc)}\\
&=\frac{vol(S^q)}{2k_D^2}\int_{\calm_d}\ d^dx\sqrt{-\tg}\biggl(r^d-\calv(\si)\biggr)\,,
\end{split}
\eqlabel{sefftot}
\end{equation}
with
\begin{equation}
\begin{split}
\calv&=\calv_{bulk}+\calv_{(loc)}\\
&=-(q-1)\ q\ \si^{2(2-d-q)/q}
+\frac 12 b^2 \si^{2-2d}+k_D^2\frac{2  T_d N}{vol(S^q)\ \si^d}\,.
\end{split}
\eqlabel{pot}
\end{equation}
It is easy to verify that  effective action \eqref{sefftot} equations of motion 
\begin{equation}
r^d_{\mu\nu}-\frac 12 r^d\ \tg_{\mu\nu}+\ft 12 \calv\ \tg_{\mu\nu}=0\,,
\eqlabel{e1}
\end{equation}
\begin{equation}
\frac{\del\calv}{\del\si}=0\,,
\eqlabel{effeom}
\end{equation}
are equivalent to $D$-dimensional equations of motion \eqref{munu}, \eqref{transf}.

\subsection{Incorrect $d$-dimensional effective description}
In this section we illustrate the pitfalls of using incomplete $D$-dimensional 
equations of motion to deduce $d$-dimensional effective action. This approach 
was used in existing literature, and has led to quite unexpected conclusions concerning 
consistency of KK reductions in the type IIB flux compactifications.  

For a constant modulus $\si$ we expect the effective action to be of the type  
\eqref{sefftot}, which we rewrite one more time as 
\begin{equation}
S^{eff}=\frac{vol(S^q)}{2k_D^2}\int_{\calm_d}\ d^dx\sqrt{-\tg}\biggl(r^d-\tcalv(\si)\biggr)\,.
\eqlabel{sdeff2}
\end{equation}
Einstein equations from \eqref{sdeff2} imply that 
\begin{equation}
\begin{split}
r^d_{\mu\nu}-\frac 12 r^d\ \tg_{\mu\nu}+\ft 12 \tcalv\ \tg_{\mu\nu}&=0\,,\\
\frac 12\ \tg_{\mu\nu}\biggl(\tcalv-(d-2)\LL_d\biggr)&=0\,,
\end{split}
\eqlabel{wrong}
\end{equation}
where in the second line we used the fact that $\calm_d$ must be an Einstein manifold.
With \eqref{wrong}, it is tempting to use $D$-dimensional equation of motion 
\eqref{munu} to conclude 
\begin{equation}
\begin{split}
\tcalv&=(d-2) \LL_d\\
&=-\frac{b^2(q-1)(d-2)}{2(d+q-2)}\si^{2(1-d)}-k_D^2\frac{T_d N}{ vol(S^q)\
\si^{d}}\ \frac{(d-2)(q-2)}{d+q-2}\,.
\end{split}
\eqlabel{wrongv}
\end{equation}
Potential $\tcalv$ \eqref{wrongv} derived in this way disagrees with  
\eqref{pot}, and is manifestly incorrect. 
For one reason, in the absence of any defects ($N=0$), $\tcalv$ implies 
that $\si$ can not be stabilized. This conclusion differs from exact $D$-dimensional 
analysis.

\section{KK reduction of type IIB supergravity with localized sources}
We now  implement the computations of the toy model in the previous 
section in type IIB supergravity compactifications on a large six-dimensional manifold 
$\tcalm_6$ with fluxes and localized 3-brane sources. 
Specifically, we obtain the correct potential for the overall 
K\"ahler modulus of $\tcalm_6$.
As the analysis are rather technical, 
we begin by stating the main conclusion. 
For the effective action obtained by integrating out the $\tcalm_6$ coordinates from the ten dimensional 
type IIB supergravity action with branes,
its equations of motion are equivalent to the ten dimensional equations of motion. 
To elucidate this agreement, one has to use the {\it full} set 
of type IIB equations of motion. This agreement is lost however is one 
evaluates the full effective action on the equations of motion for 
a subset of the supergravity modes, on-shell condition for which depends 
on the K\"ahler modulus. Such dependence is known to occur for the 
on-shell condition on the ten dimensional metric warp factors \cite{gkp}, 
and, as we argue below, also arises for the 3-form flux equations of motion.

As a byproduct of the analysis, we point out that for the nonvanishing 
four dimensional cosmological constant, the complex structure moduli of $\tcalm_6$
can not be fixed at a point where the 3-form fluxes are imaginary-self-dual 
(ISD) \cite{btalk}. The K\"ahler modulus potential derived 
here can be useful in study the backreaction of non-ISD fluxes. 
We hope to return to this problem in the 
future.

\subsection{Conventions}
Consider a static\footnote{\ie, there 
are no ``rolling'' moduli.} type IIB string theory 
compactification on a large compact manifold $\calm_6$ with fluxes.
We assume the supergravity approximation is valid.
In particular, we assume that all the relevant moduli 
are   fixed at large values so that one can safely 
ignore $\a'$ corrections. 

The low-energy effective action is decomposed into 
two parts
\begin{equation}
S_{eff}=S_{IIB}+S_{sources}\,,
\eqlabel{ac3}
\end{equation} 
where 
\begin{equation}
\begin{split}
S_{IIB}=\frac{1}{2k_{10}^2}\int_{\calm_{10}}&\ \biggl(
R_{10}\wedge \star 1 -\frac 12 d\Phi\wedge \star d\Phi
-e^{2\Phi}dC_{(0)}\wedge\star dC_{(0)}
-\frac 12 e^{-\Phi} H_3\wedge\star H_3\\
&  -\frac 12 
e^{\Phi} F_3\wedge\star F_3
 -\frac 14 F_5\wedge \star F_5-\frac 12 C_{(4)}\wedge H_3\wedge F_3\biggr)
\end{split}
\eqlabel{ssugra}
\end{equation}
is the Einstein frame action of 
type IIB supergravity, and  $S_{sources}$ is the 
effective action of the localized sources (branes wrapping various cycles 
of the compactification manifold).  
We take the ten dimensional space-time $\calm_{10}$ to be a direct warped product of 
a four dimensional spacetime $\calm_4$ and a six dimensional compact manifold $\tcalm_6$.
The Einstein frame 
metric is 
\begin{equation}
\begin{split}
ds_{10}^2=&\hg_{MN}d\xi^M d\xi^N\\
=&e^{2A(y)-6 u}\ 
g_{\mu\nu}dx^{\mu}dx^{\nu} + e^{-2A(y)+2u}\ 
\tg_{mn}(y)dy^mdy^n\,,
\end{split}
\eqlabel{km10}
\end{equation}  
where $g_{\mu\nu}$ and $\tg_{mn}$ is a  metric on $\calm_4$ and $\tcalm_6$ correspondingly, 
$A(y)$ is a warp factor, 
and $u$ defines the overall volume modulus of $\tcalm_6$ 
\begin{equation}
\si\equiv \Im\r\equiv e^{4u}\,.
\eqlabel{defr}
\end{equation}    
As we review later, the choice of relative warpings by $e^u$ in \eqref{km10} is required
to decouple the fluctuations of $\r(x)$ from the four dimensional metric fluctuations 
\cite{dg}. Additionally, there are  nontrivial 5-form $F_5$
and  3-form  
\begin{equation}
G_{3}=F_3-\tau H_3
\eqlabel{k3form}
\end{equation}
fluxes.
In \eqref{k3form} $\tau$ is the type IIB axiodilaton.
$S_{sources}$ in \eqref{ac3} provides local sources (in addition to 
those in $S_{IIB}$) for the Einstein equations, Maxwell and dilaton 
equations. For example, for a $p$-brane wrapped on a $(p-3)$-cycle
$\Sigma$ of the compactification manifold $\tilde{\calm}_6$
we have
\begin{equation}
S_{sources}=-T_p\int_{\calm_4\times\Sigma}d^{p+1}\xi \sqrt{-g}
+\mu_p\int_{\calm_4\times\Sigma}C_{p+1}\,,
\eqlabel{3brane}
\end{equation}
with 
\begin{equation}
T_p=|\mu_p|e^{(p-3)\Phi/4}\,.
\eqlabel{tension}
\end{equation}
We introduce the  stress tensor $T^{loc}_{MN}$ 
for the  localized sources in \eqref{ac3}
\begin{equation}
T^{loc}_{MN}=-\ft{2}{\sqrt{-\hg}} 
\ft{\dd (S_{sources})}{\dd \hg^{MN}}\,. 
\eqlabel{tloc}
\end{equation} 
In what follows we assume that the only localized sources are $D3$, $\bD3$ branes, 
and orientifold 3-planes $O3$.

We use a mostly positive convention for the signature $(-+\cdots +)$
and take $\epsilon_{1\cdots10}=+1$.  The bosonic type  IIB equations
consist of the following \cite{schwarz83}:

\noindent $\bullet$\quad  The Einstein equations:
\begin{equation}
R_{MN}=T^{(1)}_{MN}+T^{(3)}_{MN}+T^{(5)}_{MN}+k_{10}^2\left(
T_{MN}^{loc}-\ft 18 \hat{g}_{MN} T^{loc}\right)\,,
\eqlabel{tenein}
\end{equation}
where the energy momentum tensors of the dilaton/axion field, $\calb$,
the three index antisymmetric tensor field, $\calf_{(3)}$, and the self-dual
five-index tensor field, $\calf_{(5)}$, are given by
\begin{equation}
T^{(1)}_{MN}= P_MP_N{}^*+P_NP_M{}^*\,,
\eqlabel{enmomP}
\end{equation}
\begin{equation}
T^{(3)}_{MN}=
       \frac 18(G^{PQ}{}_MG^*_{PQN}+G^{*PQ}{}_MG_{PQN}-
        \frac 16 g_{MN} G^{PQR}G^*_{PQR})\,,
\eqlabel{enmomG}
\end{equation}
and
\begin{equation}
T^{(5)}_{MN}= \frac 16 \calf^{PQRS}{}_M\calf_{PQRSN}\,.
\eqlabel{enmomF}
\end{equation}
In the unitary gauge, $\calb$ is a complex scalar field, and
\begin{equation}
P_M= f^2\partial_M \calb\,,\qquad Q_M= f^2\,{\rm Im}\,(
\calb\partial_M\calb^*)\,,
\eqlabel{defofPQ}
\end{equation}
where
\begin{equation}
f= \frac{1}{ (1-\calb \calb^*)^{1/2}}\,,
\eqlabel{defoff}
\end{equation}
while the antisymmetric tensor field $G$ is given by
\begin{equation}
G= f(\calf_{(3)}-\calb \calf_{(3)}^*)\,.
\eqlabel{defofG}
\end{equation}

\noindent
$\bullet$\quad The Maxwell equations:
\begin{equation}
(\nabla^P-i Q^P) G_{MNP}= P^P G^*_{MNP}-\frac 23\,i\,\calf_{MNPQR}
G^{PQR}\,.
\eqlabel{tenmaxwell}
\end{equation}

\noindent
$\bullet$\quad The dilaton equation:
\begin{equation}
(\nabla^M -2 i Q^M) P_M= -\frac{1}{ 24} G^{PQR}G_{PQR}\,.
\eqlabel{tengsq}
\end{equation}

\noindent
$\bullet$\quad The self-dual equation:
\begin{equation}
\calf_{(5)}= \star \calf_{(5)}\,.
\eqlabel{tenself}
\end{equation}

\noindent
In addition, $\calf_{(3)}$ and $\calf_{(5)}$ satisfy Bianchi identities which
follow from the definition of the field strengths in terms of their
potentials:
\begin{equation}
\begin{split}
\calf_{(3)}&= dA_{(2)}\,,\\
\calf_{(5)}&= dA_{(4)}-{\frac 18}\,{\rm Im}( A_{(2)}\wedge
\calf_{(3)}^*)\,.
\end{split}
\eqlabel{defpotth}
\end{equation}
We would like to relate fluxes  $\calf_{(3)},\ \calf_{(5)}$, 
and the unitary dilaton $\calb$ to the appropriate 
quantities of \eqref{ssugra}. This has been done, among many other places, 
in  \cite{bpp}:
\begin{equation}
\begin{split}
\tau\equiv C_{(0)}+i e^{-\Phi}&=i\frac{1+\calb}{1-\calb}\,,\\
A_{(2)}&=C_{(2)}+i B_{(2)}\,,\\
A_{(4)}&=\ft 14 \left(C_{(4)}+\ft 12  B_{(2)}\wedge C_{(2)}\right)\,,
\end{split}
\eqlabel{conv}
\end{equation}
with
\begin{equation} 
F_3=dC_{(2)},\qquad H_3=dB_{(2)}\,.
\eqlabel{fh}
\end{equation}

\subsection{Ten dimensional  equations of motion}
We begin with a more general  metric ansatz (in Einstein frame) 
\begin{equation}
\begin{split}
ds_{10}^2&=\om_1^2(y)\ \ww_1^2(x)\ ds^2_{\calm_4}(x)+\om_2^2(y)\ \ww_2^2(x)\  
ds^2_{\tilde{\calm}_6}(y)\\
&=\om_1^2(y)\ \ww_1^2(x)\ g_{\mu\nu}(x) dx^{\mu}dx^{\nu} 
+\om_2^2(y)\ \ww_2^2(x)\ \tg_{mn}(y) dy^{m}dy^{n}\,,
\end{split}
\eqlabel{10metricg}
\end{equation}
where $\calm_4$ is taken to be an Einstein manifold, \ie,
\begin{equation}
r_{\mu\nu}^{(4)}(x)=\LL\ g_{\mu\nu}(x)\,,
\eqlabel{4}
\end{equation}
and $\tilde{\calm}_6$ is a six dimensional compactification manifold. 
Additionally we assume that all the stringy matter and   the
localized sources depend 
on $y$ only. The rigorous way to say this,  is that 
we require our compactification to preserve (anti-)de-Sitter 
or Poincare invariance. For the 5-form $\calf_5$ we assume 
\begin{equation}
\calf_5=\left(1+\star\right) \left[d\ww\wedge vol_{\calm_4}\right]\,,  
\eqlabel{5form}
\end{equation}
where $vol_{\calm_4}$ is the volume form on $\calm_4$.
The basic idea is to allow $\LL$ in \eqref{4} to ``dynamically 
adjust'' (letting it be either positive, zero  or negative) 
depending on the fluxes and the geometry of the 
{\it compact } $\tilde{\calm}_6$. 
We should say that the equations we obtain are equivalent 
(when they can be compared) to those presented in \cite{gkp}. 
To elucidate  the agreement with \cite{gkp},
some redefinitions are necessary.
For example, the $G_{MNP}$ we are using is related to the 
$G_3$ flux \eqref{k3form} of \cite{gkp} as 
\begin{equation}
G_3=\frac {1}{f (1-\calb)}\ G\,.
\eqlabel{matchgkp}
\end{equation} 
In what follows all the index contractions are done with 
the unwarped metrics $g_{\mu\nu},\ \tg_{mn}$. Also 
we use $r^{(4)}_{\mu\nu}$ to be the Ricci tensor
of $\calm_4$, and  $r^{(6)}_{mn}$ for the Ricci tensor 
of $\tilde{\calm}_6$. Capital $R_{\mu\nu},\ R_{mn}$ 
are reserved for the Ricci components of the full ten dimensional 
warped metric \eqref{10metricg}. The absence of indexes implies 
that they are contracted.
After straightforward  computations we find
\begin{equation}
\begin{split}
R_{\mu\nu}=&r^{(4)}_{\mu\nu}+2\biggl(2\ww_1^{-2}\n_\mu\ww_1\n_\nu\ww_1\
-\ww_1^{-1}\n_\mu\n_\nu\ww_1\biggr)\\
&+6\biggl(\ww_1^{-1}\ww_2^{-1}\biggl[\n_\mu\ww_1\n_\nu\ww_2+
\n_\mu\ww_2\n_\nu\ww_1\biggr]-\ww_2^{-1}\n_\mu\n_\nu\ww_2\biggr)\\
&-g_{\mu\nu}\biggl(\ww_1^{-1}\n^2\ww_1+\ww_1^{-2}\left(\n\ww_1
\right)^2+6\ww_1^{-1}\ww_2^{-1}\n\ww_1\n\ww_2+\om_2^{-2}\ww_2^{-2}
\om_1\ww_1^2\tn^2\om_1\\
&+3\om_2^{-2}\ww_2^{-2}\ww_1^2\left(\tn\om_1\right)^2
+4\om_2^{-3}\ww_2^{-2}\om_1\ww_1^2\tn\om_1\tn\om_2
\biggr)\,,
\end{split}
\eqlabel{ricci4}
\end{equation}
where 
\begin{equation}
\n\equiv \n_x\,,\qquad \tn\equiv \n_y\,, 
\end{equation} 
and
\begin{equation}
\begin{split} 
R_{mn}=&r^{(6)}_{mn}+4\biggl(2\om_2^{-2}\tn_m\om_2\tn_n\om_2\
-\om_2^{-1}\tn_m\tn_n\om_2\biggr)\\
&+4\biggl(\om_1^{-1}\om_2^{-1}\biggl[\tn_m\om_1\tn_n\om_2+
\tn_m\om_2\tn_n\om_1\biggr]-\om_1^{-1}\tn_m\tn_n\om_1\biggr)\\
&-\tg_{mn}\biggl(\om_2^{-1}\tn^2\om_2+3 \om_2^{-2}\left(\tn\om_2
\right)^2+4\om_1^{-1}\om_2^{-1}\tn\om_1\tn\om_2\\
&+\om_1^{-2}\ww_1^{-2}\om_2^2\ww_2\n^2\ww_2
+5\om_1^{-2}\ww_1^{-2}\om_2^2\left(\n\ww_2\right)^2
+2\ww_1^{-3}\ww_2\om_1^{-2}\om_2^{2}\n\ww_1\n\ww_2
\biggr)\,.
\end{split}
\eqlabel{ricci6}
\end{equation}
Also, $R_{\mu m}\ne 0$, but we will not need its explicit 
form in what follows.
The  choice of warp factors as in \eqref{km10},   
\begin{equation}
\begin{split}
&\om_1=e^{A(y)}\,,\qquad \om_2=e^{-A(y)}\,,\\
&\ww_1=e^{-3u(x)}\,,\qquad \ww_2=e^{u(x)}\,,
\end{split}
\eqlabel{gaugeg}
\end{equation}
substantially simplifies computations.
With \eqref{ricci4}-\eqref{gaugeg} we find
\begin{equation}
\begin{split}
R_{10}\sqrt{-\hg}=&e^{-4A}\sqrt{\tg}\sqrt{-g}\biggl(r^{(4)}
+6\n^2u-24 \left(\n u\right)^2\biggr)\\
&+e^{-8u}\sqrt{\tg}\sqrt{-g}\biggl(r^{(6)}+2\tn^2 A-8\left(
\tn A\right)^2\biggr)\,.
\end{split}
\eqlabel{EHkk}
\end{equation}
Note that terms containing 
$\n^2 u$ and $\tn^2 A$ are total derivatives and 
can be dropped from the effective action.
From \eqref{EHkk}, the four dimensional gravitational constant $k_4$  in terms 
of ten dimensional gravitational constant $k_{10}$ is given by
\begin{equation}
\frac{1}{k_4^2}=\frac{1}{k_{10}^2}\ \int_{\tcalm_6}\sqrt{\tg} e^{-4A}\,.
\eqlabel{k4}
\end{equation}
Thus, as in \cite{gkp,dg}, we get from \eqref{EHkk} the kinetic 
term for the $\Im \r\equiv e^{4u}$ modulus
\begin{equation}
\frac{1}{k_4^2}\int_{\calm_4}d^4x\sqrt{-g}\biggl\{-3\frac{\n\bar{\r}\n\r}
{|\r-\bar{\r}|^2}\biggr\}\,,
\eqlabel{rmod}
\end{equation}
corresponding to the K\"ahler potential
\begin{equation}
\calk_\r=-3\ln(-i (\r-\bar{\r}))\,.
\eqlabel{kr}
\end{equation}
From now on we assume that the modulus $\r$ is constant 
over $\calm_4$.  
Note that with $\r$ being constant 
\begin{equation}
R_{\mu m}=0\,.
\eqlabel{offdiag}
\end{equation}
From \eqref{3brane}, \eqref{tloc} localized stress tensor contribution
from 3-branes, $T^{3-brane}_{MN}$, is 
\begin{equation}
T_{\mu\nu}^{3-brane}=-T_3 e^{2A}\left(\Im\r\right)^{-3/2}\ g_{\mu\nu}\
\r_3^{loc},\qquad T_{mn}^{3-brane}=0\,,
\eqlabel{stressloc}
\end{equation}
where $\r_3$ is the number density of 3-branes: 
\begin{equation}
\#\ (3-branes)
=\int_{\calm_6} d^6 y\ \sqrt{\tg} e^{-6A}\left(\Im\r\right)^{3/2}
\r_3^{loc}\,.
\eqlabel{numberdensity}
\end{equation}
For a single 3-brane localized at $y=y_0$ we have
\begin{equation}
\r_3^{loc}=e^{6A}\left(\Im\r\right)^{-3/2}\ \frac{\dd^6(y-y_0)}{\sqrt{\tg}}\,.
\eqlabel{r31}
\end{equation}
Since $O3$ tension is $-\frac 14 T_3$,   
contribution to the localized stress tensor from a single $O3$ plane will be
minus a quarter of that of a $D3$ brane. To treat orientifold planes and 3-branes on the same footing
we write the full localized stress tensor as  
\begin{equation}
T_{\mu\nu}^{loc}=- e^{8A}\left(\Im\r\right)^{-3}\ g_{\mu\nu}\
\sum_{sources} T_i\ \frac{\dd^6(y-y_i)}{\sqrt{\tg}}\,,\qquad T_{mn}^{loc}=0\,,
\eqlabel{tlocf}
\end{equation}
where the summation is over all  sources  localized at $y=\{y_i\}$ and having 
tension 
\begin{equation}
\begin{split}
T_i\equiv T_3\,,\qquad &{\rm source}_i\equiv \{D3,\bD3\}\,,\\
T_i\equiv -\frac 14 T_3\,,\qquad &{\rm source}_i\equiv \{O3\}\,.
\end{split}
\eqlabel{tidef}
\end{equation}

The simplest Einstein equation is for the $\{\mu\nu\}$ components  
\begin{equation}
\begin{split}
\LL  -\left(\Im \r\right)^{-2}
e^{4A}\tn^2 A =&-\left(\Im \r\right)^{-3}\
\ft {1}{48} G_{mnp}\bG^{mnp} e^{8A}
-\left(\Im \r\right)^{4}\ 4 (\tn\ww)^2 e^{-4 A}\\
&-\frac 12 k_{10}^2 e^{8A}\left(\Im\r\right)^{-3}\ \sum_{sources} T_i\ \frac{\dd^6(y-y_i)}{\sqrt{\tg}}\,,
\end{split}
\eqlabel{emunu}
\end{equation}  
or 
\begin{equation}
\begin{split}
\tn^2 e^{4A}=&\left(\Im \r\right)^{-1}\ 
\ft {1}{12} e^{8A} G\bG +e^{-4 A} \biggl(
16\ \left(\Im\r\right)^6
\left(\tn\ww\right)^2+\left(\tn e^{4A}\right)^2 
\biggr)\\
&+\left(\Im\r\right)^2\left(
4\LL+2k_{10}^2 e^{8A}\left(\Im\r\right)^{-3}
\ \sum_{sources} T_i\ \frac{\dd^6(y-y_i)}{\sqrt{\tg}}
\right)\,.
\end{split}
\eqlabel{e4a}
\end{equation}
Note the shorthand notation for $G\bG$.

The next simplest equation is for the 5-form Bianchi identity
\begin{equation}
\begin{split}
\left(\Im\r\right)^3\ \tn^2\ww=&
\left(\Im\r\right)^3\
2 e^{-4A}\tn\ww \tn e^{4A}+\left(\Im\r\right)^{-1}\
\ft {i}{48} e^{8A} G\star_6\bG\\
&-\ft 12 k_{10}^2e^{8A} \left(\Im\r\right)^{-1}\  
\sum_{sources} q_i |T_i|\ \frac{\dd^6(y-y_i)}{\sqrt{\tg}}\,,
\end{split}
\eqlabel{bianchi}
\end{equation}
where $\star_6$ is defined on  manifold $\tilde{\calm}_6$,
and $q_i=\pm1$ is a  charge of a localized source at $y=y_i$. 
Note that above normalization of $q_i$ is correct for  $O3$ planes as well:
their `physical' charge is $\left(-\frac 14\right)$ of  a $D3$ brane charge.

From \eqref{e4a} and \eqref{bianchi} we find an
important constraint ---
a simple modification of  eq.~(2.30) of \cite{gkp}
\begin{equation}
\begin{split}
\tn^2\left(\left(\Im\r\right)^3 4\ww+e^{4A}\right)=&e^{-4 A}\left(\tn\left[
\left(\Im\r\right)^3
4\ww+e^{4A}\right]\right)^2 +
\left(\Im\r\right)^{-1}{\textstyle\frac{1}{24}} e^{8A} |i G
+\star_6 G|^2\\+\left(\Im\r\right)^2\ 4\LL
&+\left(\Im\r\right)^{-1}\ 
2 k_{10}^2 e^{8A}\ \sum_{sources} \left(T_i-q_i|T_i|\right) \ \frac{\dd^6(y-y_i)}{\sqrt{\tg}}\,.
\end{split}
\eqlabel{const}
\end{equation}
To proceed with the remaining equations of motion 
we separate imaginary (anti-) self dual parts of $G$:
\begin{equation}
\begin{split}
G^+\equiv\ft 12 G-\ft i2 \star_6 G\,,\\
G^-\equiv\ft 12 G+\ft i2 \star_6 G\,,
\end{split}
\eqlabel{iasd}
\end{equation}
where 
\begin{equation}
\star_6 G^\pm=\pm i G^\pm\,.
\eqlabel{check}
\end{equation}
It will also be convenient to introduce a 3-form 
$\call$
\begin{equation}
\call=e^{4A}\star_6 G-4i\ww\left(\Im\r\right)^3 G
=-i(e^{4A}+4\omega\left(\Im\r\right)^3)G^- + i 
(e^{4A}-4\omega\left(\Im\r\right)^3)G^+\,.
\eqlabel{ldef}
\end{equation} 

The remaining Einstein equations are
\begin{equation}
\begin{split}
r^{(6)}_{mn}=&\ft 12 e^{-8A}\biggl(\tn_m e^{4A}\tn_n e^{4A}-16 
\left(\Im\r\right)^6\ \tn_m\ww\tn_n\ww\biggr)
-\LL\ \left(\Im\r\right)^2\ e^{-4 A} \tg_{mn} +T_{mn}^{(1)}\\
&+\left(\Im\r\right)^{-1}\
\ft 14 e^{4A} \biggl(G^{+}\ _{pqm}\bG^{-pq}\ _n+G^{-}\
_{pqm}\bG^{+pq}\  
_n
\biggr)\,,
\end{split}
\eqlabel{emn}
\end{equation}
where $T^{(1)}_{mn}$ is the standard dilaton/axion stress 
tensor
\begin{equation}
T_{mn}^{(1)}=\ft 14 \frac{\tn_m\tau\tn_n\bar{\tau}+
\tn_n\tau\tn_m\bar{\tau}}{(\Im{\tau})^2}\,.
\eqlabel{t1}
\end{equation}
The 3-form Maxwell equations \eqref{tenmaxwell} are reduced to
\begin{equation}
0=d\call+f^2\biggl(\bar{\call}\wedge d\calb+\ft 12 \call\wedge \left(
\calb d\bar{\calb}-\bar{\calb}d\calb\right)\biggr)\,.
\eqlabel{max}
\end{equation}
Finally, the dilaton equation is
\begin{equation}
f^2 \tn^2\calb+2f^4\bar{\calb}(\tn\calb)^2=-\ft{1}{12} e^{6A} G^+G^-\,.
\eqlabel{dil}
\end{equation}

Integrating both sides of \eqref{const} over compact manifold 
$\tilde{\calm}_6$  we find
\begin{equation}
\begin{split}
\LL=&\biggl\{-\frac{1}{4(\Im\r)^2\ vol_{\tilde{\calm_6}}}\
\int_{\tilde{\calm}_6}d^6y \sqrt{\tg} \biggl(
e^{-4 A}\left(\tn\left[
\left(\Im\r\right)^3
4\ww+e^{4A}\right]\right)^2 \\
&+\left(\Im\r\right)^{-1}{\textstyle\frac{1}{6}} e^{8A} |G^+|^2
\biggr)
\ \biggr\}\\
+& \biggl\{-\frac{k_{10}^2} {2(\Im\r)^3\ vol_{\tilde{\calm_6}}}
\sum_{sources} e^{8A(y_i)} \left(T_i-q_i|T_i|\right)
\ \biggr\}\,,
\end{split}
\eqlabel{llf}
\end{equation}
where we separated contributions to the four dimensional cosmological 
constant coming from supergravity modes,  and the localized
sources.
Also, 
\begin{equation}
\ vol_{\tilde{\calm_6}}=\int_{\tilde{\calm}_6}d^6y\sqrt{\tg}\,.
\eqlabel{v6}
\end{equation}

We would like to conclude the section with several comments concerning solutions 
of equations of motion derived here. 
\nxt First of all, notice that the only contribution to the four dimensional 
cosmological constant from localized sources comes from $\bD3$ branes. 
Both $D3$ branes and $O3$ planes saturate the BPS-like bound of GKP (which can be thought 
of as a generalized strong energy condition for charged defects in ten dimensions), and do not contribute to $\LL$. 
$\bD3$ branes satisfy this bound, and thus provide negative contribution to $\LL$.
As the result,  supergravity plus localized 
3-branes and orientifold 3-planes flux compactifications can never lead to  
four dimensional de Sitter. This result is straightforward to generalize to all extended objects in 
string theory satisfying the BPS-like condition of GKP. Note that the neutral defects in the toy model 
of section 2 satisfy the GKP bound. 
   
To circumvent this no-go theorem for de-Sitter compactifications, one possibility is to consider 
flux compactifications with localized sources
that violate BPS-like condition of GKP. These include $\overline{O3}$ planes and orientifold 5-planes, 
$O5$.  Currently no construction of this type is known.  Another possibility is to relax the 
condition that all moduli of the compactification manifold are stabilized. There are interesting 
cosmological solutions with rolling moduli \cite{tw}. However, to our knowledge, 
no nonsingular solution of this type (in supergravity with branes) exists. 
Most promising appear to be the framework of KKLT \cite{kklt}, where one first employs nonperturbative
string corrections to stabilize all moduli with supersymmetric $AdS_4$ vacuum, and further lifts
it to a $dS_4$ with  supersymmetry breaking branes. Here, it is important to better understand
effects of the supersymmetry breaking, in particular the role of non-ISD fluxes
that can be induced by these supersymmetry breaking effects.
\nxt The 3-form Maxwell equation \eqref{max} allow for a solution
\begin{equation}
\call=0\,.
\eqlabel{maxsol}
\end{equation} 
We do not know whether \eqref{maxsol} gives the most general solution of the 3-form Maxwell equation \eqref{max},
though we suspect this is the case. 
The string theory flux backgrounds constraint by \eqref{maxsol} include those  discussed in \cite{gkp,b1}.
This is  the class of compactifications we concentrate on here.  
Notice  that \eqref{maxsol} does not imply  that the 3-form flux is ISD,
\ie, 
\begin{equation}
G^+=0\,.
\eqlabel{isdw}
\end{equation}
In fact, this is simply inconsistent once $\LL\ne 0$. 
Indeed, given \eqref{isdw}, eqs.~\eqref{ldef}, \eqref{maxsol} then imply that  
$e^{4A}+4\ww\left(\Im\r\right)^3=0$. This last condition,
along with $G^+=0$, implies from \eqref{const}
\begin{equation}
0=\left(\Im\r\right)^2\ 4\LL
+\left(\Im\r\right)^{-1}\ 
2 k_{10}^2 e^{8A} \sum_{sources}
\left(T_i-q_i |T_i|\right)\ \frac{\dd^6(y-y_i)}{\sqrt{\tg}}\,.
\eqlabel{constISD}
\end{equation}
Clearly, \eqref{constISD} can not be satisfied locally, given that 
cosmological constant  $\LL$ is {\it uniform}
over $\tcalm_6$. 
\nxt It is tempting to use \eqref{llf}, and follow the arguments of Sec.~2.3, to conclude 
that the K\"ahler modulus potential $\tcalv(\r)$ is 
\begin{equation}
\begin{split}
\tcalv(\r)=&2 \LL\\
=&\biggl\{-\frac{1}{2(\Im\r)^2\ vol_{\tilde{\calm_6}}}\
\int_{\tilde{\calm}_6}d^6y \sqrt{\tg} \biggl(
e^{-4 A}\left(\tn\left[
\left(\Im\r\right)^3
4\ww+e^{4A}\right]\right)^2 \\
&+\left(\Im\r\right)^{-1}{\textstyle\frac{1}{6}} e^{8A} |G^+|^2
\biggr)
\ \biggr\}\\
&+ \biggl\{-\frac{k_{10}^2 }{(\Im\r)^3\ vol_{\tilde{\calm_6}}}
\sum_{sources} e^{8A(y_i)}\left(T_i-q_i|T_i|\right)
\ \biggr\}\,.
\end{split}
\eqlabel{wrongviib}
\end{equation}
This is incorrect, as the use of $\tcalv$ in the effective action 
analogous to \eqref{sdeff2} implies that $\rho$ can be off-shell; because of this,
this K\"ahler potential can never be deduces from on-shell quantities, 
like equations of motion. The exactly soluble toy model of section 2 
explicitly demonstrates this point. 
\nxt
We would like to present  a simple expression for $\LL$, following from the full set of ten dimensional
equations of motion 
\begin{equation}
\LL=-\frac{k_4^2}{2}\sum_{sources} \biggl(T_i e^{4 A(y_i)}\left(\Im\r\right)^{-3}+4 q_i |T_i|\ \ww(y_i)\biggr)\,.
\eqlabel{corrll}
\end{equation}   
The derivation of \eqref{corrll} is delegated to Appendix.
It would be nice to find a simpler derivation of \eqref{corrll};
in particular it's extension for more general F-theory flux compactifications. 

Notice that \eqref{corrll} suggests that the four dimensional cosmological constant is sensitive 
to exactly how $\bD3$ branes are distributed\footnote{As explained in KKLT, $\bD3$ branes are driven 
to the end of the 'Klebanov-Strassler throat' of $\tcalm_6$, where the warp factor $e^A$ is locally minimized. }
between different Klebanov-Strassler-like 'throats' of 
$\tcalm_6$. This opens a possibility of interesting cosmological transitions where in evolution 
to global minimum $\LL$, the Universe experiences metastable vacua with {\it different}
cosmological constants\footnote{Related issues have been discussed in \cite{af}.}. 
We plan to study these transitions in the future.

\subsection{Four dimensional effective description}
In this section we derive the four dimensional effective description of supergravity and localized sources 
flux compactifications, implementing the arguments of Sec.~2.2. We find that equations of motion 
derived from this effective description are equivalent to a subset of ten dimensional equations of motion.  
The K\"ahler modulus potential is given by Eq.~\eqref{potfinal}.

Recall that \eqref{ssugra}
\begin{equation}
\begin{split}
S_{IIB}=&\frac{1}{2k_{10}^2}\int_{\calm_{10}}d^{10}\xi\ \sqrt{-\hg}
\biggl\{R_{10}-\frac{\del_M\tau\del^M\bar{\tau}}{2(\Im\tau)^2}
-\frac{G\cdot \overline{C}}{12}
-\frac{F^2_{(5)}}{4\cdot 5!}
\biggr\}\\
&+\frac{1}{8i k_{10}^2}\int_{\calm_{10}}\ C_{(4)}\wedge G
\wedge \overline{G}\,.
\end{split}
\eqlabel{iibfull}
\end{equation}
Let's evaluate the 
ten dimensional effective action \eqref{ac3} for the metric ansatz \eqref{10metricg},
\eqref{gaugeg}. 
We rewrite \eqref{ac3} as 
\begin{equation}
S_{eff}^{(4)}=S_{IIB}^{(4)}+S_{sources}^{(4)}\,,
\eqlabel{eff4}
\end{equation}
where 
\begin{equation}
\begin{split}
S_{IIB}^{(4)}=&\frac{1}{2k_{10}^2}\int_{\calm_{4}}d^4x\sqrt{-g}
\int_{\tcalm_6}d^6y\sqrt{\tg}\ \biggl(
r^{(4)} e^{-4A(y)} \\
&+\frac{1}{\left(\Im\r\right)^2}\biggl[r^{(6)}(y)-8\left(\tn 
A(y)\right)^2-\frac{\left(\tn\tau(y)\right)^2}{2(\Im\tau)^2}
\biggr]
-\frac{1}{12 (\Im\r)^3} e^{4A(y)} G\cdot \bar{G}(y)\\
&+8\ e^{-8 A(y)} \left(\Im\r\right)^4\left(\tn\ww(y)\right)^2
\biggr)+CS\,,
\end{split}
\eqlabel{iib40}
\end{equation}
where we used \eqref{conv} and \eqref{5form}\footnote{One has to be careful 
with evaluation of the action of the self-dual 5-form. A correct prescription to do this 
was explained in \cite{dg}.}.
$CS$ denotes the type IIB supergravity 
Chern-Simons term, see \eqref{iibfull}. 
Explicitly,
\begin{equation}
\begin{split}
CS=&\frac{1}{8i k_{10}^2}\int_{\calm_{10}}\ 8\ww\ vol_{\calm_4}\wedge
G\wedge \bar{G}\\
=&\frac{1}{2k_{10}^2}\int_{\calm_{4}}d^4x\sqrt{-g}
\int_{\tcalm_6}d^6y\sqrt{\tg}\  \biggl(
\frac{i\ww}{3} G\cdot \star_6\bar{G}
\biggr)\,.
\end{split}
\eqlabel{cs}
\end{equation}
Note an extra factor of $2$ for the $C_{(4)}$ in \eqref{cs} --- again, this is required 
for proper dimensional reduction of the self-dual 5-form, \cite{dg}. 
Thus \eqref{iib40} reads
\begin{equation}
\begin{split}
S_{IIB}^{(4)}=&\frac{1}{2k_{10}^2}\int_{\calm_{4}}d^4x\sqrt{-g}
\int_{\tcalm_6}d^6y\sqrt{\tg}\ \biggl(
r^{(4)} e^{-4A(y)}\\
&+\frac{1}{\left(\Im\r\right)^2}\biggl[r^{(6)}(y)-8\left(\tn 
A(y)\right)^2-\frac{\left(\tn\tau(y)\right)^2}{2(\Im\tau)^2}
\biggr]
+8\ e^{-8 A(y)} \left(\Im\r\right)^4\left(\tn\ww(y)\right)^2
\\
&+\frac{1}{12 (\Im\r)^3}  G\cdot \star_6\bar{\call}(y)
\biggr)\,.
\end{split}
\eqlabel{iib4}
\end{equation}
Finally, the dimensional reduction of the localized sources (again,
only $D3,\bD3$ branes and $O3$ planes) \eqref{3brane} reads 
\begin{equation}
S_{sources}^{(4)}=-\sum_{sources}\biggl
(T_i e^{4 A(y_i)}\left(\Im\r\right)^{-3}+4q_i|T_i|\ww(y_i)\biggr)\int_{\calm_4}d^4x\sqrt{-g}\,.
\eqlabel{effsource}
\end{equation}
Now, rewriting   effective action \eqref{eff4} as  
\begin{equation}
S_{eff}^{(4)}=\frac{1}{2k_4^2}\int_{\calm_4}d^4x\sqrt{-g}\biggl(r^{(4)}-\calv(\rho)\biggr)\,,
\eqlabel{4final}
\end{equation}
we conclude from \eqref{iib4} and \eqref{effsource}
\begin{equation}
\begin{split}
\calv(\r)=&-\left(\Im\r\right)^{-2}\ \frac{k_{4}^2}{k_{10}^2}\ \int_{\tcalm_6}d^6y\sqrt{\tg}
\left[r^{(6)}(y)-8\left(\tn 
A(y)\right)^2-\frac{\left(\tn\tau(y)\right)^2}{2(\Im\tau)^2}
\right]\\
&-\left(\Im\r\right)^{4}\ \frac{k_{4}^2}{k_{10}^2}\ \int_{\tcalm_6}d^6y\sqrt{\tg}\left[
8\ e^{-8 A(y)}\left(\tn\ww(y)\right)^2
\right]\\
&+\left(\Im\r\right)^{-3}\ \frac{k_{4}^2}{k_{10}^2}\ \int_{\tcalm_6}d^6y\sqrt{\tg}\left[
\frac{1}{12 }  e^{4A(y)}\ G\cdot \bar{G}(y)
\right]\\
&-\frac{k_{4}^2}{k_{10}^2}\ \int_{\tcalm_6}d^6y\sqrt{\tg}\left[
\frac{i\ww}{3} G\cdot\star_6\bar{G}
\right]\\
&+2 k_4^2\sum_{sources}\biggl
(T_i e^{4 A(y_i)}\left(\Im\r\right)^{-3}+4q_i|T_i|\ww(y_i)\biggr)\,,
\end{split}
\eqlabel{potfinal}
\end{equation}
where we organized contributions according to various scaling with $\left(\Im\r\right)\equiv \si$:
\begin{equation}
\calv(\r)\equiv\calv(\si)\equiv\si^{-2}\ V_{-2}+\si^{4}\ V_4+\si^{-3}\ V_{-3}+V_0+V_{sources}\,,
\eqlabel{potsh}
\end{equation}
where $\{V_{-2}, V_4,V_{-3}, V_0,V_{soruces}\}$ can be read off comparing 
\eqref{potfinal} and \eqref{potsh}.
We would like to claim that \eqref{potfinal} is the correct off-shell potential for $\r$.

Consider Einstein equations derived from \eqref{4final}.
They imply that $\calm_4$ is an Einstein manifold 
\begin{equation}
r^{(4)}_{\mu\nu}=\LL^{(4)}\ g_{\mu\nu}\,,
\eqlabel{ein44}
\end{equation}
with 
\begin{equation}
\LL^{(4)}=\frac 12\calv(\si)\,.
\eqlabel{ll4}
\end{equation}
We demonstrate now  that $\LL^{(4)}$ is actually equivalent to 
ten dimensional expression \eqref{corrll}, once ten dimensional on-shell condition, \ie, equations of motion, is used.
Indeed, because of \eqref{maxsol}, 
\begin{equation}
\si^{-3}\ V_{-3}+V_0=0\,.
\eqlabel{ch1}
\end{equation}
Now the trace of \eqref{emn} implies 
\begin{equation}
\left(\Im\r\right)^{-2}
\left[r^{(6)}-8\left(\tn 
A\right)^2-\frac{\left(\tn\tau\right)^2}{2(\Im\tau)^2}
\right]+\left(\Im\r\right)^4\left[
8\ e^{-8 A}\left(\tn\ww\right)^2
\right]=-6\LL e^{-4 A}\,,
\eqlabel{traceMN}
\end{equation}
or, equivalently, 
\begin{equation}
\si^{-2}\ V_{-2}+\si^4\ V_4=6\LL\,. 
\eqlabel{tracee}
\end{equation}
Using \eqref{corrll}, we conclude that 
\begin{equation}
V_{sources}=-4\LL\,.
\eqlabel{vsfin}
\end{equation} 
Given \eqref{ch1}, \eqref{tracee}, \eqref{vsfin} we conclude from \eqref{potsh}
\begin{equation}
\calv(\si)=2\LL\,,
\eqlabel{potshc}
\end{equation}
which along with \eqref{ll4} gives
\begin{equation}
\LL^{(4)}=\LL\,.
\eqlabel{ll4ll}
\end{equation}

In the rest of this section we show that the equation of motion for the K\"ahler modulus $\si$ 
\begin{equation}
\frac{d\calv(\si)}{d\si}\equiv \calv'=0\,,
\eqlabel{vp}
\end{equation}
is satisfied as well.
From \eqref{potsh},
\begin{equation}
\si\ \calv'=-2 \si^{-2}\ V_{-2}+4 \si^4\ V_4-3\si^{-3}\ V_{-3}-6k_4^2\sum_{sources} T_ie^{4A(y_i)} \si^{-3}\,,
\eqlabel{dirv}
\end{equation}
where the last term comes from the derivative of $V_{sources}$.
Using \eqref{tracee}, Eq.~\eqref{dirv} becomes
\begin{equation}
\si\ \calv'=-12 \LL+ 6 \si^4\ V_4-3\si^{-3}\ V_{-3}-6k_4^2\sum_{sources} T_ie^{4A(y_i)} \si^{-3}\,.
\eqlabel{dirv1}
\end{equation}
Notice from \eqref{llap2}
\begin{equation}
\LL=-\frac{k_4^2}{2}\sum_{sources} T_ie^{4A(y_i)}\ \si^{-3}-\frac 14 \si^{-3}\ V_{-3}+\frac 12 \si^4\ V_4\,, 
\eqlabel{ch2}
\end{equation}
thus Eq.~\eqref{dirv1} becomes
\begin{equation}
\begin{split}
\si\ \calv'=&-12 \LL+ \biggl(
12\LL+6k_4^2\sum_{sources} T_ie^{4A(y_i)} \si^{-3}
\biggr)-6k_4^2\sum_{sources} T_ie^{4A(y_i)} \si^{-3}
\\
=&0\,.
\end{split}
\eqlabel{dirv2}
\end{equation}

\subsection{Concluding remarks}
We would like to emphasize that to obtain agreement 
between ten dimensional and four dimensional effective descriptions 
of string theory flux compactifications\footnote{In the supergravity 
approximation with localized D-branes sources.} one has to use 
the full potential \eqref{potfinal}. In particular, it is 
incorrect to evaluate compact supergravity modes on their equations 
of motion in the expression for $\calv(\si)$. 
This was explicitly demonstrated in Sec.~2.3 for the case of the 
exactly soluble toy model. 

Similar phenomenon can be observed in the current context. 
Indeed, ten dimensional equations of motion allow for a class 
of solutions with  
$\call=0$, \eqref{maxsol}. The latter is equivalent to 
the relation \eqref{ch1}. However, we can not impose
\eqref{ch1} at the level of the potential $\calv$ \eqref{potfinal},
so that 
\begin{equation}
\calv(\si)\to\hat{\calv}(\si)=\si^{-2}\ V_{-2}+\si^{4}\ V_4+V_{sources}\,.
\eqlabel{wrongpots}
\end{equation}   
It is easy to see that though $\hat{\calv}$ leads to the same value of the 
four dimensional cosmological constant, the equation of motion for the K\"ahler modulus $\si$ 
is inconsistent with ten dimensional equations of motion, unless $G=0$. 
Again, the problem appears because $\calv(\si)$ is an off-shell quantity,
and its partial evaluation on-shell, and subsequent treatment of the 
result as off-shell quantity is inconsistent. From a  slightly different angle, 
the  inconsistency occurred because equations of motion for the fluxes 
{\it depend} on the overall size modulus $\sigma$ (a component of the 
ten dimensional metric), thus 
\begin{equation}
\biggl(\left[\frac{d\calv(\sigma)}{d\sigma}\right]'\biggr)\bigg|_{EOM\{fluxes\}}\ \ne\ 
\frac{d}{d\sigma}\left[\calv(\sigma)\bigg|_{EOM\{fluxes\}}\right]\,, 
\eqlabel{eom}
\end{equation}
since 
\begin{equation}
\frac{d}{d\sigma}\biggl[{EOM\{fluxes\}}\biggr]\ \ne\ 0\,,
\eqlabel{eom1}
\end{equation}
where  $EOM\{fluxes\}$ stands for the fluxes equations of motion.
Note, that it is not only on-shell fluxes that  depend on the K\"ahler modulus.
As explained in  \cite{ds2}, the contribution to the four dimensional 
K\"ahler modulus effective  potential from the $\bD3$ brane, $\delta V$, does not scale 
as $\sigma^{-3}$ as assumed in \cite{kklt}. Rather, because the 
on-shell warp factor $e^{4A}\sim \sigma$ \cite{gkp}, one finds (see 
Eq.~(5.14) of \cite{ds2})
\begin{equation}
\begin{split}
\delta V&\sim \frac{1}{\sigma^3}e^{4A}\sim \frac{1}{\sigma^2}\,,\\
 &\equiv \frac{D}{(2\sigma)^2}\,.
\end{split}
\eqlabel{truesc}
\end{equation}

Finally, in deriving the potential \eqref{potfinal}, we did not assume that 
the 3-from fluxes are ISD, \ie, $G^+=0$, or that $\tcalm_6$ is Ricci flat,
also we did not assume  supersymmetry. 
Thus potential \eqref{potfinal} can be useful to study effects of 
supersymmetry breaking and the backreaction of non-ISD fluxes in 
KKLT-like models.


\section*{Acknowledgments}
I would like  to thank Shanta de Alwis, Steve Giddings,
 Shamit Kachru,  Renata Kallosh, Jim Liu, Rob Myers,  
Joe Polchinski, Radu Roiban, Christian Romelsberger, Sav Sethi, Henry Tye,  
and Johannes Walcher for numerous discussions, correspondence and comments. 
I would like to thank Aspen Center for Physics
for hospitality where part of this work was done, and 
acknowledge the 
partial support by the U.S. Department of Energy while at the University 
of Michigan, where this work started. 
Research at the Perimeter Institute is supported in part by funds from NSERC of 
Canada.


\section*{Appendix}
Here we derive Eq.~\eqref{corrll}. First, multiply both sides of \eqref{emunu} by $e^{-4 A}\sqrt{\tg}$, 
and integrate over $\tcalm_6$. Using definition \eqref{k4}, and the fact that 
\begin{equation}
\int_{\tcalm_6}d^6y \sqrt{\tg}\ \tn^2 A=0\,, 
\eqlabel{ap1}
\end{equation}
we conclude 
\begin{equation}
\begin{split}
\LL=&-\frac{k_4^2}{2}\sum_{sources} T_i e^{4 A(y_i)}\left(\Im\r\right)^{-3}\\
&-\frac{k_4^2}{k_{10}^2}\int_{\tcalm_6}d^6y\sqrt{\tg}\biggl(
\left(\Im \r\right)^{-3}\
\ft {1}{48} G_{mnp}\bG^{mnp} e^{4A}
+\left(\Im \r\right)^{4}\ 4 (\tn\ww)^2 e^{-8 A}
\biggr)\,.
\end{split}
\eqlabel{llap2}
\end{equation}
Let's multiply both sides of the Bianchi identity  \eqref{bianchi} by 
$e^{-8A}\ww \sqrt{\tg}$ and integrate over $\tcalm_6$. We find
\begin{equation}
\begin{split}
&\frac{k_4^2}{k_{10}^2}\int_{\tcalm_6}d^6y\sqrt{\tg}\biggl(
\left(\Im\r\right)^{4}e^{-8A} \ww\left(\tn^2\ww-8\tn\ww\tn A \right)-
\ft{i}{48}\ww\ G\star_6\bar{G} 
\biggr)\\
&=\\
&-\frac{k_4^2}{2}\sum_{sources} q_i |T_i|\ \ww(y_i)\,.
\end{split}
\eqlabel{intb}
\end{equation}
Given \eqref{llap2} and \eqref{intb}, Eq.~\eqref{corrll} follows, 
provided 
\begin{equation}
\begin{split}
\cali\equiv &4 \left(\Im\r\right)^4
e^{-8A}\biggl(\ww\tn^2\ww-8 \ww\tn\ww\tn A+\left(\tn\ww\right)^2\biggr)\\
&+\biggl(-\ft{i}{12} \ww\ G\star_6\bar{G}+\left(\Im\r\right)^{-3}\ft{1}{48} e^{4A}\ G\bar{G}\biggr)
\end{split}
\eqlabel{ape3}
\end{equation}
integrates to zero over $\tcalm_6$,
\begin{equation}
\int_{\tcalm_6}d^6y\sqrt{\tg}\ \cali=0\,.
\eqlabel{intzero}
\end{equation}
But \eqref{intzero} is trivial once we notice that 
\begin{equation}
\cali=4\left(\Im\r\right)^4\tn\left(e^{-8 A}\ww\tn\ww\right)-\ft{1}{48}
\left(\Im\r\right)^{-3}
\ G\star_6\bar{\call}\,,
\eqlabel{i2}
\end{equation}
and recall that the 3-form Maxwell equation \eqref{max} implies \eqref{maxsol}.

\end{document}